\documentclass[a4paper,review]{elsarticle}
\pdfoutput=1 
\journal{Journal of Futures Markets}

\usepackage{amsmath,amssymb,graphicx,hyperref,natbib,multirow,ifthen,setspace,xcolor}
\usepackage[left=1in,right=1in,top=1.4in,bottom=1.2in]{geometry}

 \biboptions{authoryear} 
\hypersetup{
	pdftitle={A Black-Scholes user's guide to the Bachelier model},
	pdfauthor={Jaehyuk Choi, Minsuk Kwak, Chyng Wen Tee, Yumeng Wang},
	pdfkeywords={Bachelier model, Black-Scholes model, Displaced diffusion model, Normal model},
	colorlinks=true,
	linkcolor=red,
	citecolor=blue,
	urlcolor=blue,
	bookmarksnumbered=true,
	pdfstartview=
}

\newcommand{\qtext}[2][\quad]{#1\text{#2}#1}

\newcommand{\sigman}{\sigma_\textsc{n}}
\newcommand{\sigmad}{\sigma_\textsc{d}}
\newcommand{\sigmabs}{\sigma_\textsc{bs}}
\newcommand{\sigmacev}{\sigma_\textsc{cev}}
\newcommand{\norm}{\textsc{n}}
\newcommand{\cev}{\textsc{cev}}
\newcommand{\bs}{\textsc{bs}}
\newcommand{\disp}{\textsc{d}}
\newcommand{\sv}{\textsc{sv}}
\newcommand{\betac}{\beta_\ast}
\newcommand{\betacpow}[1]{\beta_\ast^{#1}}
\newcommand{\ncx}{{\chi^2}}
\newcommand{\vov}{\nu}
\newcommand{\rhoc}{\rho_\ast}
\newcommand{\distequal}{\,{\buildrel d \over =}\,}
\newcommand{\Zdrift}[1][\mu]{\ifthenelse{\equal{#1}{0}}{Z}{Z^{[#1]}}}
\newcommand{\asinh}{\mathrm{asinh}}

\newcommand{\atanh}{\mathrm{atanh}}
\newcommand{\Call}[2]{C_\textsc{#1}^\textsc{#2}}
\newcommand{\Put}[2]{P_\textsc{#1}^\textsc{#2}}

\begin{document}
\begin{frontmatter}

\title{A Black--Scholes user's guide to the Bachelier model}

\date{14 January, 2022}

\author[phbs]{Jaehyuk Choi\corref{corrauthor}}
\ead{jaehyuk@phbs.pku.edu.cn}

\author[hufs]{Minsuk Kwak}
\ead{mkwak@hufs.ac.kr}

\author[smu]{Chyng Wen Tee}
\ead{cwtee@smu.edu.sg}

\author[jtb]{Yumeng Wang}
\ead{yumeng_wang@bankcomm.com}

\cortext[corrauthor]{Corresponding author \textit{Tel:} +86-755-2603-0568, \textit{Address:} Rm 755, Peking University HSBC Business School, University Town, Nanshan, Shenzhen 518055, China}

\address[phbs]{Peking University HSBC Business School, Shenzhen, China}
\address[hufs]{Department of Mathematics, Hankuk University of Foreign Studies, Yongin, Republic of Korea} 
\address[smu]{Lee Kong Chian School of Business, Singapore Management University, Singapore}
\address[jtb]{Bank of Communications, Shanghai, China}

\begin{abstract} 
To cope with the negative oil futures price caused by the COVID--19 recession, global commodity futures exchanges temporarily switched the option model from Black--Scholes to Bachelier in 2020. This study reviews the literature on Bachelier's pioneering option pricing model and summarizes the practical results on volatility conversion, risk management, stochastic volatility, and barrier options pricing to facilitate the model transition. In particular, using the displaced Black--Scholes model as a model family with the Black--Scholes and Bachelier models as special cases, we not only connect the two models but also present a continuous spectrum of model choices.
\end{abstract}
\begin{keyword}
	Bachelier model, Black--Scholes model, Displaced diffusion model, Normal model\\
	\textit{JEL Classification}: G10, G13
\end{keyword}
\end{frontmatter}

\section{Introduction} \noindent
Louis Bachelier pioneered an option pricing model in his Ph.D. thesis~\citep{bachelier1900}, marking the birth of mathematical finance. He offered the first analysis of the mathematical properties of Brownian motion (BM) to model the stochastic change in stock prices, and this preceded the work of \citet{einstein1905bm} by five years. His analysis also precursors what is now known as the efficient market hypothesis~\citep{schachermayer2008close}. See \citet{sullivan1991louis} for Bachelier's contribution to financial economics and \citet{courtault2000louis} for a review of his life and achievements.

Owing to the celebrated Black--Scholes (BS) model~\citep{blackscholes1973,merton1973theory} and the fact that the arithmetic BM allows negative asset prices, the Bachelier model has been forgotten as a part of history until recently. Ironically, the model gained attention again in the twenty-first century because it can deal with negative asset prices, which was considered its limitation. The negative interest rates observed in some developed countries
after the 2008 global financial crisis forced fixed-income trading desks to reconsider their option pricing models. The spread of COVID--19 led to lockdowns worldwide and an extremely sharp drop in the global demand for oil. Consequently, in April 2020, the price of oil futures contracts became sharply negative for the first time in history. In response, the Chicago Mercantile Exchange (CME) and Intercontinental Exchange (ICE) temporarily changed their models for oil and natural gas options from the BS to the Bachelier model until August 2020 to handle the negative prices~\citep{cme2020bachelier,cme2020naturalgas,cme2020transition,ice2020negative}.

In fact, the attention on the Bachelier model dates back to the pre-2008 crisis era, even when the fear of negative prices was negligible. The Bachelier model has been widely used at least in the fixed income markets---swaptions are quoted and risk-managed by Bachelier volatility.\footnote{The Bachelier model is also called the \textit{normal} model as it assumes a normal distribution of the asset price. As such, the term ``\textit{normal} volatility'' is more widely used than ``Bachelier volatility.'' However, we use the term ``Bachelier volatility'' throughout the paper for consistency.} This is because the proportionality between the daily changes in and the level of the interest rate---a key assumption of the BS model---is empirically weak. Consequently, the log-normal distribution cannot accurately describe the interest rate dynamics~\citep{levin2002interest,levin2004rate,ho2003interest}.
The ICE's MOVE index, also referred to as ``the VIX for bonds,'' is the weighted Bachelier volatility from short-term US Treasury Bond options.\footnote{The acronym MOVE originally stands for Merrill-Lynch Option Volatility Estimate.}
Other than in fixed income markets, the Bachelier model was naturally adopted when the underlying price can assume negative values. For example, it has been used for spread options (i.e., the option on the price difference of two assets)~\citep{poitras1998spread} and year-on-year inflation~\citep{kenyon2008inflation}.

Despite the recent surge of interest in the Bachelier model, it is still difficult to find comprehensive references on the model.\footnote{For example, the Wikipedia entry for ``Bachelier model'' (\url{https://en.wikipedia.org/wiki/Bachelier_model}) was created on April 21, 2020, presumably after the CME announcement. Accessed as of January 14, 2022, the entry does not even contain the option price formula, Eq.~\eqref{eq:bach}.} The academic literature on the Bachelier model is scarce, or at best, scattered over different papers, each addressing only certain aspects of the model, and many of which are unpublished preprints. We aim to fill this gap by reviewing the Bachelier model in a way that benefits both researchers and practitioners. The attempt to review the Bachelier model is certainly not new, but existing reviews focus on the historical perspectives of Bachelier and his model~\citep{sullivan1991louis,courtault2000louis,schachermayer2008close}.\footnote{\citet{brooks2017arithmetic} is an exception. While they review various strengths of the Bachelier model over the BS model, they interpret the Bachelier negative price under the Bachelier model as the nonzero probability of equity price hitting zero, which differs from our assumption that the price can go freely negative.} 
Instead, we review the Bachelier model in a modern context as an alternative option pricing model to BS, as in the case with the recent model change.

We briefly summarize the key aspects of this review. First, we aim to provide actionable assistance to practitioners who are considering switching from (or using in parallel with) the BS model. For example, we provide an analytic conversion formula between the different model volatilities (Section~\ref{ssec:conversion}). We explain the delta hedging under the Bachelier model in terms of the vega-rotated delta under the Black--Scholes model (Section~\ref{ssec:backbone}). Second, we review the Bachelier and BS models as two special cases of more general model families such as the displaced Black--Scholes (DBS) or stochastic-alpha-beta-rho (SABR) model. We show that one can easily obtain the results for the Bachelier model by continuously transforming those of the BS model. This framework also offers a spectrum of model choices in terms of the volatility skew, leverage effect, and allowance of negative prices rather than a binary choice between the Bachelier and BS models. Lastly, this paper also offers a novel contribution to the literature, although it is a review paper. We improve the accuracy of the analytical volatility conversion to the Bachelier volatility (Section~\ref{ssec:conversion}) and present the barrier option pricing formulas under the Bachelier model (Section~\ref{ssec:barrier}). Our review of the DBS model is new in the literature, to the best of our knowledge, although it is easily generalized from the BS model. Overall, we aim to provide a definitive one-stop reference for an overview of known results on the Bachelier model, its application as a reporting and pricing model, and its connection to the BS and DBS models.

The remainder of this paper is organized as follows. Section \ref{sec:model} introduces the Bachelier model. Section~\ref{sec:related_models} reviews the DBS and SABR models. In Section \ref{sec:vol}, we focus on volatility-related topics such as implied volatility inversion and conversion between the models. Section~\ref{sec:greeks} discusses the Greeks and hedging, and Section~\ref{sec:sv} reviews the stochastic Bachelier volatility model. Section~\ref{sec:exotic} covers the pricing of exotic claims under the Bachelier model, and we offer our conclusions in Section~\ref{sec:conc}.

\section{Bachelier model} \label{sec:model} \noindent
\subsection{Bachelier and BS models} \noindent
The Bachelier model assumes that the $T$-forward price of an asset at time $t$, $F_t$, follows an arithmetic BM with volatility $\sigman$,
$$ dF_t = \sigman \, dW_t 
$$
where $W_t$ is a standard BM under the $T$-forward measure. The undiscounted price of a call option with strike price $K$ and time-to-maturity $T$ under the Bachelier model is \footnote{The put option price under the Bachelier model is
$$ P_\norm(K) = (K - F_0) N(-d_\norm) + \sigman\sqrt{T} \, n(d_\norm)
$$}
\begin{equation} \label{eq:bach}
C_\norm(K) = (F_0 - K) N(d_\norm) + \sigman\sqrt{T} \, n(d_\norm)
\qtext{for} d_\norm = \frac{F_0 - K}{\sigman \sqrt{T}},
\end{equation}
where $n(z)$ and $N(z)$ are the probability density function (PDF) and cumulative distribution function (CDF), respectively, of the standard normal distribution. With the continuously compounded interest rate, $r$, and convenience yield of the asset, $q$, we can express the option price by the spot price $S_0 = e^{(q-r)T}\,F_0$ instead of the forward price $F_0$; the discounted option price is $e^{-rT} C_\norm(K)$. As Bachelier himself noted, the option price (both put and call) at the money (ATM), $K=F_0$, is simplified and the volatility is easily inverted as~\citep[\S 2.1]{schachermayer2008close}
\begin{equation} \label{eq:atm}
	C_\norm(F_0) = \sigman \sqrt{\frac{T}{2\pi}} \qtext{and} \sigman = C_\norm(F_0) \sqrt{\frac{2\pi}{T}}.
\end{equation}

The Bachelier formula also holds under a weaker assumption. As long as the asset price at maturity, $F_T$, is normally distributed with mean $\mu(F_T)$ and standard deviation $sd(F_T)$, we can derive the undiscounted call option price with a minor modification:
\begin{equation} \label{eq:bach_gen}
C_\norm(K) = sd(F_T) \left( d_\norm N(d_\norm) + n(d_\norm) \right)
\qtext{for} d_\norm = \frac{\mu(F_T) - K}{sd(F_T)}.
\end{equation}
This generalized formula is helpful in pricing a basket options and Asian options, which we discuss further in Section~\ref{sec:exotic}.

In contrast, the BS model~\citep{blackscholes1973,black1976} assumes a geometric BM with volatility $\sigmabs$,
$$ \frac{dF_t}{F_t} = \sigmabs\, dW_t.
$$
The corresponding undiscounted call option price is well known as the \citet{black1976} formula\footnote{The put option price under the BS model is 
	$$P_\bs(K) = K\, N(-d_2) - F_0\,N(-d_1).$$}:
\begin{equation} \label{eq:bsm}
	C_\bs(K) = F_0\,N(d_1) - K\, N(d_2)
	\qtext{for}
	d_{1,2} = \frac{\log (F_0/K)}{\sigmabs\sqrt{T}} \pm \frac{\sigmabs\sqrt{T}}{2}.
\end{equation}
We can obtain the \citet{blackscholes1973} formula easily by substituting $F_0 = e^{(r-q)T}\,S_0$ and discounting the premium by $e^{-rT}$.

Volatility has different meanings in each model; while the BS volatility $\sigmabs$ measures the relative change in $F_t$, the Bachelier volatility $\sigman$ measures the absolute change in $F_t$. The relation, $\sigman = \sigmabs F_0$, ensures that the dynamics between the two models behave similarly within a short time interval, and that the two models yield a similar ATM option price, 
$$ C_\bs(F_0) \approx C_\norm(F_0) \approx 0.4\, \sigmabs F_0 \sqrt{T}.$$ 
Many Wall-Street options traders use this approximation as a back-of-the-envelope calculation for the ATM BS option price.

\subsection{Alternative specification of the Bachelier model} \noindent
Other studies~\citep{brooks2017arithmetic} apply the arithmetic BM to the spot price $S_t$ instead of the forward price $F_t$. In that case, the dynamics is given by an Ornstein--Uhlenbeck process:
$$ dS_t = (r-q) S_t dt + \sigman' \, dW_t. $$
It is worth noting the difference between this and Eq.~\eqref{eq:bach}. From $F_t = e^{(T-t)(r-q)}\,S_t$, we can show that the equivalent dynamics on $F_t$ are 
$$ dF_t = e^{(r-q)(T-t)} \sigman' \, dW_t.$$
Therefore, the main difference in the alternative approach is that volatility is increasing or decreasing exponentially. It can still take advantage of the Bachelier formula because $F_T$ is normally distributed. By integrating the variance, the standard deviation of $F_T$ is
$$ \quad sd(F_T) = \sigman' \sqrt{\frac{e^{2(r-q)T} - 1}{2(r-q)}} \qquad \left( = \sigman'\sqrt{T} \qtext{if} r=q \right),
$$
which we can plug into the generalized Bachelier formula in Eq.~\eqref{eq:bach_gen}. Since Eq.~\eqref{eq:bach} with $\sigman = sd(F_T)/\sqrt{T}$ will produce the same option price, we view the two approaches as equivalent when pricing vanilla options.
Since we aim to cover options traded on futures exchanges, we choose to work with Eq.~\eqref{eq:bach}, where the volatility of the forward price is constant. See \citet{takehara2010new} for a further discussions of other specifications.

\section{Models generalizing the Bachelier and BS models}\label{sec:related_models} \noindent
In this section, we review the models that bridge the Bachelier and BS models as two special cases: 
the DBS and SABR models. We will show that the Bachelier model is a limit case of the DBS model, making it possible to continuously transform the DBS model into either the Bachelier or the BS model, and vice versa. The analysis of the SABR model is useful for the convenient conversion of the volatilities between different models. 

\subsection{Displaced BS model} \noindent
The DBS model is a popular way to adjust the BS model to allow negative asset prices and negative volatility skew without sacrificing the analytical tractability of the BS model~\citep{rubinstein1983displaced,joshi2003lmmsv}. While there are various specifications, we present the DBS model with volatility $\sigmad$ as follows:\footnote{Our DBS model specification with two parameters, $\beta$ and $A$, is general enough to include the two alternative specifications,
	$$ D(F_t) = F_t + A \qtext{or} D(F_t) = \beta F_t + (1-\beta) F_0. $$
	Compared to the first, ours helps to clarify the Bachelier model in the $\beta\downarrow 0$ limit. We also intentionally avoid the second because the dependency on $F_0$ in $D(F_t)$ may causes unintended confusion in computing delta, the partial derivative with respect to $F_0$. The DBS model with $D(F_t) = \beta F_t + (1-\beta)F_0$ and $\sigmad= F_0^{\beta-1}\sigmacev$ also serves as an approximation to the constant-elasticity-of-variance (CEV) model~\citep{svoboda2009disp}.}
\begin{equation} \label{eq:D}
\frac{dF_t}{D(F_t)} = \sigmad\; dW_t \qtext{where} D(F_t) = \beta\,F_t + (1-\beta)A.
\end{equation}
Under the DBS model, the \textit{displaced} variable, $D(F_t)$, rather than $F_t$, follows a geometric BM with volatility $\sigmad$, and the model can handle negative underlying prices with the lower bound $F_t>-(1-\beta)A/\beta$. The final asset price, $F_T$ is accordingly 
$$ F_T = \left(F_0+\frac{1-\beta}{\beta}A\right) \exp\left(\beta\sigmad W_T-\frac{\beta^2\sigmad^2T}{2}\right) - \frac{1-\beta}{\beta} A,
$$
and the call option price is
\begin{equation} \label{eq:dbs0}
	C_\disp(K)= \frac{D(F_0) N(d_{1\disp}) - D(K) N(d_{2\disp})}{\beta}
	\qtext{for}
	d_{1\disp,2\disp} = \frac{\log\left(D(F_0)/D(K)\right)}{\beta\sigmad\sqrt{T}} \pm \frac{\beta\sigmad\sqrt{T}}{2}.
\end{equation}
In other words, we can re-use the Black formula in Eq.~\eqref{eq:bsm} by replacing $F_0$, $K$, $\sigmabs$, and $C_\bs$ with $D(F_0)$, $D(K)$, $\beta\sigmad$, and $\beta C_\disp$, respectively. We can extend other analytical results for the BS model to the DBS model with little difficulty; see Section~\ref{sec:greeks} for the Greeks. 

From the model dynamics in Eq.~\eqref{eq:D}, it is clear that the BS model is a special case of the DBS model with $\beta=1$ and $\sigmad=\sigmabs$, and that the Bachelier model is another with $\beta=0$ and $\sigmad=\sigman/A$. The BS option price is trivially reduced from that of the DBS model. However, some effort is required to obtain the Bachelier option price from the $\beta\downarrow 0$ limit of the DBS option price.
For small $\beta$, we have the following approximations: 
\begin{gather*}
\log\left(\frac{D(F_0)}{D(K)}\right) = \frac{\beta (F_0-K)}{(1-\beta)A} \left(1+\frac{\beta(F_0+K)}{2(1-\beta)A}\right) + O(\beta^2), \\
d_{1\disp,2\disp} = \frac{ F_0-K}{(1-\beta)A \sigmad\sqrt{T}} \left(1+\frac{\beta(F_0+K)}{2(1-\beta)A}\right)
\pm \frac{\beta\sigmad\sqrt{T}}{2} + O(\beta).
\end{gather*}
Now, we show that the DBS price converges to the Bachelier price as $\beta\downarrow 0$:
\begin{align*}
	C_\disp(K) &= \frac{D(F_0)-D(K)}{\beta} N(d_{2\disp}) + \frac{D(F_0)}{\beta}\left(N(d_{1\disp}) - N(d_{2\disp})\right) \\
	& = (F_0-K) N(d_{2\disp}) + \frac{D(F_0)}{\beta} (d_{1\disp}-d_{2\disp})\, n(d_{1\disp}) + O(\beta) \\
	& \rightarrow (F_0-K) N(d_\norm) + \sigman\sqrt{T}\, n(d_\norm) = C_\norm(K) \qtext{with} \sigman = A\sigmad.
\end{align*}
Understanding the Bachelier model as the $\beta\downarrow 0$ limit of the DBS model is very helpful throughout this paper, as we can use this method to verify many results. See Section~\ref{sec:greeks} for a discussion of the Greeks and Section \ref{ssec:barrier} for details on Barrier option pricing. 

\begin{figure} 
	\caption{\label{fig:skew} The BS volatility skew of the Bachelier, DBS ($\beta=1/3$ and $2/3$ with $A=F_0$), and BS models. We use $F_0=1$, $T=1$, and $\sigman=\sigmad=\sigmabs=0.5$.}
	\begin{center}
		\includegraphics[width=0.55\linewidth]{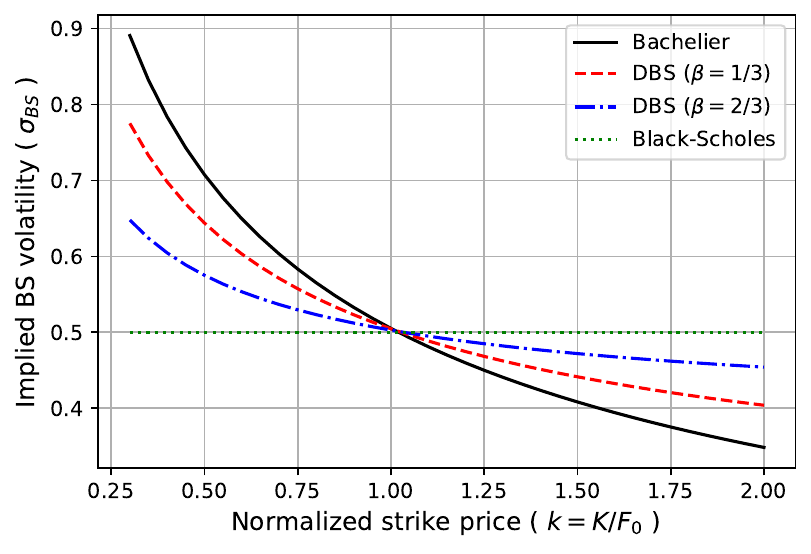}
	\end{center}
\end{figure}

We next discuss the leverage effect and negative skew of the DBS and Bachelier models. In the DBS model, we find similar dynamics (and option price) to the BS model in a small time horizon when
$$ \sigmabs\,F_0 \approx \sigmad(\beta F_0+ (1-\beta)A).
$$
For a fixed $\sigmad$, the equivalent $\sigmabs$ increases when $F_0$ decreases, achieving the leverage effect. In particular, the degree of the leverage effect strengthens as $\beta$ decreases, reaching the maximum at $\beta=0$ (i.e., the Bachelier model).\footnote{Although we consider $\beta\in[0, 1]$ in this paper, practitioners sometimes use $\beta<0$ or $\beta>1$ to achieve super-normal or super-log-normal skewness, respectively. In the case of $\beta<0$, $A$ can also be negative to ensure the inequality $F_t > -(1-\beta)A/\beta$.} To illustrate the effect, we show the BS volatility skew implied from the DBS models with a fixed $\sigmad$ but varying $\beta$ values in Figure~\ref{fig:skew}. As expected, the skew becomes more negative as $\beta$ decreases from 1 (BS) to 0 (Bachelier). Therefore, the DBS model is a simple alternative to the BS model for modeling the negative skew observed in the market, where $\beta$ is used to fit the slope. 

\subsection{SABR (and CEV) model} \label{ssec:sabr} \noindent
The SABR model~\citep{hagan2002sabr} is a stochastic volatility (SV) model given by
\begin{equation} \label{eq:sabr}
	\frac{dF_t}{F_t^\beta} = \sigma_t\, dW_t, \quad
	\frac{d \sigma_t}{\sigma_t} = \vov \, dZ_t, \qtext{and}
	dW_t\,dZ_t = \rho\, dt,
\end{equation}
where $F_t$ and $\sigma_t$ are the processes for the forward price and volatility, respectively. $\vov$ is the volatility of volatility, $\beta$ is the elasticity parameter, and $W_t$ and $Z_t$ are the standard BMs correlated by $\rho$. Thanks to the intuitive dynamics and parsimonious parametrization, the SABR model gained popularity among practitioners, and the approximate BS volatility formula~\citep[Eq.~(A.59)]{hagan2002sabr} used to price options has become an industry standard. See \citet{antonov2012advanced} for an extensive review. Although SV models are not our primary focus here, we leverage the rich academic results for the SABR model to extend the Bachelier model.

The SABR model is understood as an SV extension of the CEV model, whose dynamics are given by
\begin{equation} \label{eq:cev}
	\frac{d F_t}{F_t^\beta} = \sigmacev\, dW_t.
\end{equation}
We can write the call option price under the CEV model analytically as\footnote{The put option price under the CEV model is
	$$P_\cev(K) = K \, \bar{F}_\ncx \left(\frac{F_0^{2\betac}}{\betacpow{2}\sigmacev^2 T};\, \frac1{\betac},\frac{K^{2\betac}}{\betacpow{2}\sigmacev^2 T}\right) - F_0\, F_\ncx \left(\frac{K^{2\betac}}{\betacpow{2}\sigmacev^2 T}; \,2+\frac1{\betac},\frac{F_0^{2\betac}}{\betacpow{2}\sigmacev^2 T}\right).$$
} 
\begin{equation} \label{eq:cev_call}
	C_\cev(K) = F_0 \, \bar{F}_\ncx \left(\frac{K^{2\betac}}{\betacpow{2}\sigmacev^2 T}; \,2+\frac1{\betac},\frac{F_0^{2\betac}}{\betacpow{2}\sigmacev^2 T}\right) - K \, F_\ncx \left(\frac{F_0^{2\betac}}{\betacpow{2}\sigmacev^2 T};\, \frac1{\betac},\frac{K^{2\betac}}{\betacpow{2}\sigmacev^2 T}\right), 
\end{equation}
where $F_\ncx(\,\cdot\,;r, x_0)$ and $\bar{F}_\ncx(\,\cdot\,;r,x_0)$ are respectively the left- and right-tail CDFs of the non-central chi-squared distribution with degrees of freedom $r$ and non-centrality parameter $x_0$, and $\betac = 1-\beta$ for notational simplicity.\footnote{See \citet{larguinho2013cev} for the Greeks and a fast numerical approximation of the analytic price of the CEV model.} At first glance, the CEV model might seem to be another connection between the Bachelier $(\beta=0)$ and BS $(\beta=1)$ models, with $\beta$ serving the same role as in the DBS model. Unfortunately, this is not the case because the CEV model does not allow for negative prices at all when $\beta>0$, unlike the DBS model. In fact, one must impose an absorbing boundary explicitly at the origin for $F_t$ to be a martingale and remain arbitrage-free. The CEV option price in Eq.~\eqref{eq:cev_call} indeed imposes the probability mass absorbed at $K=0$. As such, in the $\beta\downarrow 0$ limit, Eq.~\eqref{eq:cev_call} converges to the price of the Bachelier model with the absorbing boundary at $K=0$, not to the price in Eq.~\eqref{eq:bach} without the absorbing boundary.

The SABR model also exhibits a similar mass at zero, since the SABR model generalizes the CEV model to include SV.\footnote{See \citet{yang2018survival}, \citet{gulisashvili2018mass}, and \citet{choi2021sabrcev,choi2021note} for a discussion of the mass at zero under the CEV and SABR models.}
Interestingly, the asymptotic analysis used to derive the equivalent volatility of the SABR model does not \textit{feel} the boundary as it only concerns the neighborhood of $F_0$ in a short time. Therefore, the equivalent volatility in the $\beta\downarrow 0$ limit fortuitously assumes the Bachelier model without boundary, which is our focus. Specifically, we will depend on the equivalent Bachelier volatility of the SABR model~\citep[Eq.~(14)]{hagan2014arbfree}:
\begin{equation} \label{eq:hagan-new}
\sigman(K) \approx \sigma_0 F_0^\beta H(z) \frac{k-1}{q} \left(1 + \left(
\log \left(\frac{q\,k^{\beta/2}}{k-1}\right) \frac{\alpha^2}{q^2} + \frac{\rho}{4} \frac{k^\beta-1}{k-1} \alpha\vov + \frac{2-3\rho^2}{24} \vov^2
\right)T\right),
\end{equation}
where the intermediate variables are 
\begin{gather*}
k = \frac{K}{F_0},\quad \alpha = \frac{\sigma_0}{F_0^{\betac}}, \quad
q = \int_1^k k^{-\beta}dk = 
\begin{cases}
\dfrac{k^{\betac}-1}{\betac} & \text{if}\;\; 0\le \beta < 1 \\
\;\;\log k & \text{if}\quad \beta=1
\end{cases}, \quad
z = \frac{\vov}{\alpha}q,\\
\qtext{and} H(z) = z \Big/ \log\left( \frac{\sqrt{1+2\rho z+z^2}+z+\rho}{1+\rho} \right) \quad (H(0)=1).
\end{gather*}
This approximation serves our purpose as the outcome is the Bachelier volatility, and exhibits the desired property that $\sigman(K) = \sigma_0$ for all (even negative) $K$ when $\beta=0$ and $\vov\downarrow 0$. Moreover, Eq.~\eqref{eq:hagan-new} is a more accurate approximation of the SABR model than the original HKLW formula~\citep{hagan2002sabr} that uses the equivalent BS volatility. If necessary, one can convert the Bachelier volatility obtained above the BS volatility readily via the option price. We will rely on Eq.~\eqref{eq:hagan-new} for our discussion in Sections~\ref{ssec:conversion} and \ref{sec:sv}.

\section{Volatility} \label{sec:vol} \noindent
In the discussions in the following sections, it is important to note that the Bachelier and BS models are primarily used as \textit{reporting models} to report option prices via their implied volatilities. Because implied volatilities provide a unified measure of the relative prices of options with different strike prices and maturities, traders and brokers use them to communicate option prices. In this section, we discuss the Bachelier model volatility within the context of a reporting model and derive efficient formulas to perform volatility inversion from prices and volatility conversion from the BS implied volatilities. The choice of the reporting model does not affect the price of the vanilla option because prices are determined by supply and demand in the market. The reporting model, however, affects the hedging ratio and margin requirement, which we elaborate on further in Sections~\ref{sec:greeks}.
We also highlight that the Bachelier model alone is rarely used as a \textit{pricing model} to calculate an arbitrage-free fair value, as it cannot fit the market volatility smile. The same limitation holds for the BS model. However, we discuss the Bachelier SV model in Section~\ref{sec:sv} as a possible pricing model that can fit the market volatility smile.

\subsection{Volatility inversion} \label{ssec:iv} \noindent
Efficient computation of the implied volatility for a given option premium is important. Due to the lack of analytical form, the implied volatility must be computed iteratively through a root search algorithm. Given how routine and extensive these computations are performed across the global option markets, even a minor improvement in computational efficiency can result in significant advantage in practice. Therefore, efficient volatility inversion methods has been a research topic of interest in computational finance. For the progress made in implied volatility computation under the BS model, see \citet{li2008approx,jackel2015let,stefanica2017explicit,potz2019chebyshev}.

Similar to the BS model, computing the implied Bachelier volatility for a given option price is an important task in practice. In the Bachelier model, the volatility inversion can be reduced to finding the inverse of a univariate function on $d_\norm$, which makes the problem easier than in the BS model. Taking advantage of this fact, \citet{choi2009impvol} first express the implied volatility as
\begin{equation} \label{eq:iv}
	\sigman = \sqrt\frac{\pi}{2T}\;(2 C - \theta(F_0-K))\;h(\eta) \qtext{for}
	\eta = \frac{v}{\atanh\left(v\right)} \;\;\text{and}\;\; v = \frac{|F_0 - K|}{2C - \theta(F_0-K)},
\end{equation}
where $C$ is the undiscounted price of either a call ($\theta=1$) or put ($\theta=-1$) option. On the one hand, $2C - \theta(F_0-K)$ in the denominator of $v$ indicates the straddle price; that is, the sum of the call and put option prices at the same strike from the put-call parity.\footnote{The price of a straddle option, if directly quoted in the market, can replace $2C - \theta(F_0-K)$ in the formula.} On the other hand, $|F_0-K|$ is the intrinsic value of the straddle option. Therefore, the variable $v$ is the intrinsic-to-option value ratio of straddle option, which is also a measure of the moneyness ranging from $v=0$ ($F_0=K$) to $v=1$ ($|F_0-K| \rightarrow \infty$).

Using the rational Chebyshev approximation, they obtain a very accurate approximation for $h(\eta)$:
$$ h(\eta) \approx \sqrt{\eta}\; \frac{\sum_{k=0}^{7} a_k \eta^k}{1 + \sum_{k=1}^{9} b_k \eta^k}, \quad 
$$
with the coefficients,
\begin{equation}
\begin{matrix}
a_0 &= \texttt{3.99496 16873 45134 e-1}\\
a_1 &= \texttt{2.10096 07950 68497 e+1}\\
a_2 &= \texttt{4.98034 02178 55084 e+1}\\
a_3 &= \texttt{5.98876 11026 90991 e+2}\\
a_4 &= \texttt{1.84848 96954 37094 e+3}\\
a_5 &= \texttt{6.10632 24078 67059 e+3}\\
a_6 &= \texttt{2.49341 52853 49361 e+4}\\
a_7 &= \texttt{1.26645 80513 48246 e+4}\\
\end{matrix}
\quad \quad
\begin{matrix}
b_1 &= \texttt{ 4.99053 41535 89422 e+1}\\
b_2 &= \texttt{ 3.09357 39367 43112 e+1}\\
b_3 &= \texttt{ 1.49510 50083 10999 e+3}\\
b_4 &= \texttt{ 1.32361 45378 99738 e+3}\\
b_5 &= \texttt{ 1.59891 96976 79745 e+4}\\
b_6 &= \texttt{ 2.39200 88917 20782 e+4}\\
b_7 &= \texttt{ 3.60881 71083 75034 e+3}\\
b_8 &= \texttt{-2.06771 94864 00926 e+2}\\
b_9 &= \texttt{ 1.17424 05993 06013 e+1}.
\end{matrix}
\end{equation}
The approximation is accurate for virtually all practical purposes, and can be used without further refinement. \citet{choi2009impvol} report that the error of $h(\eta)$ from the true value is in the order of $10^{-10}$ for $|d_\norm| \le 7.7$. In the near-the-money region ($|d_\norm| \le 1.46$), where most options lie, the error decreases further to the order of $10^{-13}$. This approximation almost exactly preserves the ATM inversion in Eq.~\eqref{eq:atm} as $h(1) \approx 1 - 7\times 10^{-16}$. Note that $\eta$ has a removable singularity at $v=0$. Although not explicitly mentioned in \citet{choi2009impvol}, when $v$ is close to zero, $\eta$ should be evaluated using the Taylor's expansion,
$$ \eta = \frac{1}{1+v^2/3+v^4/5+\cdots}.
$$
\citet{jackel2017normvol} and \citet{floc2016fast} also provide alternative approximation methods for the Bachelier volatility inversion.

\subsection{Volatility conversion between the BS and Bachelier models} \label{ssec:conversion} \noindent
Here, we introduce the formulas to convert between the Bachelier, BS, and DBS models. Although conversions can be performed through numerical computation via the option price, the conversion formula in this section serves as a quick approximation and provides insights on the relations between the models.

We begin with the ATM case where analytical conversion is obvious and precise. Similar to the Bachelier model, the volatility inversion under the BS and DBS models is also analytically possible ATM because the pricing formulas simplify, respectively, to~\citep[Proposition 3.2]{dimitroff2016lnvsnorm}
\begin{equation} \label{eq:atm2}
C_\bs(F_0) = F_0 \left[ 2 N\left( \frac{\sigmabs\sqrt{T}}{2}\right) -1
\right] \qtext{and}
C_\disp(F_0) = \frac{D(F_0)}{\beta} \left[ 2 N\left( \frac{\beta \sigmad\sqrt{T}}{2}\right) -1
\right].
\end{equation}
By equating the ATM prices in Eqs.~\eqref{eq:atm} and \eqref{eq:atm2}, we can convert the DBS volatility $\sigmad$ to $\sigman$ and $\sigmabs$, respectively, as
\begin{align} 
	\sigman(F_0) &= \frac{D(F_0)}{\beta} \sqrt{\frac{2\pi}{T}} \left( 2 N\left( \frac{\beta \sigmad\sqrt{T}}{2}\right) -1 \right),\label{eq:dbs2n_atm} \\ 
	\sigmabs(F_0) &= \frac{2}{\sqrt{T}}N^{-1}\left( \frac{D(F_0)}{\beta F_0} N\left( \frac{\beta \sigmad\sqrt{T}}{2}\right) - \frac{D(F_0)}{2\beta F_0} + \frac12\right). \label{eq:dbs2bs_atm}
\end{align}

\begin{figure} 
	\caption{\label{fig:vol_conv} The equivalent Bachelier volatility $\sigman(K)$ implied from the BS model with $\sigmabs=2$, $F_0=1$, and $T=1$. Among the two approximations, our approximation in Eq.~\eqref{eq:bs2n} is closer to the exact values than that of \citet{grunspan2011note} and \citet{hagan2014arbfree} in Eq.~\eqref{eq:grunspan}.}
	\begin{center}
	\includegraphics[width=0.55\linewidth]{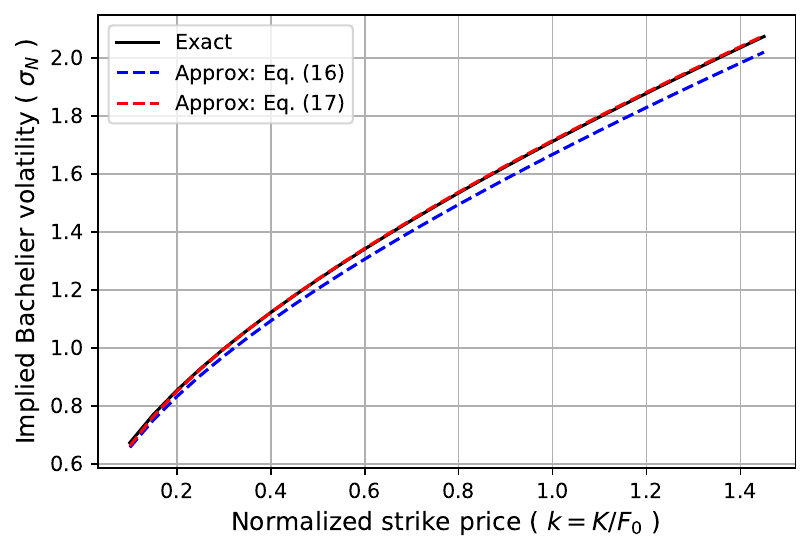}
	\end{center}
\end{figure}

For the general case of $K\neq F_0$, we first work on the conversion from $\sigmabs$ to $\sigman(K)$, which will be helpful for the transition from the BS to the Bachelier model in the oil futures case. To this end, we take advantage of the implied Bachelier volatility of the SABR model in Eq.~\eqref{eq:hagan-new}. Because the SABR model with $\beta=1$ converges to the BS model under the zero vol-of-vol limit (i.e., $\vov\downarrow 0$), Eq.~\eqref{eq:hagan-new} with $\beta=1$ and $\vov=0$ gives a conversion from the BS to Bachelier model volatility:
\begin{equation} \label{eq:grunspan}
\sigman(K) \approx \sigmabs F_0 \frac{k-1}{\log k} \left(1 - 
\log \left(\frac{k-1}{\sqrt{k}\log k}\right) \frac{\sigmabs^2 T}{\log^2 k}\right) \qtext{for} k=\frac{K}{F_0}.
\end{equation}
\citet[Corollary~2]{grunspan2011note} obtain the same result. However, we make two improvements to this formula. First, we simplify the two occurrences of $(k-1)/\log k$ to remove the singularity at $k=1$. Using the expansions near $k=1$,
$$\frac{k - 1}{\sqrt{k}} = 2\sinh\left(\,\log \sqrt{k}\right) = \log k \left(1+ \frac{\log^2 k}{24} +\frac{\log^4 k}{1920} + \cdots \right),
$$
we make the following two approximations:
$$ \frac{k-1}{\log k} \approx \sqrt{k}\left(1+\frac{\log^2 k}{24}\right) \qtext{and}
\log \left(\frac{k-1}{\sqrt{k}\log k}\right) \frac{1}{\log^2 k} \approx \frac{1}{24}.
$$
Second, we replace the $O(T)$ correction term in the form of $(1-aT)$ with $1/(1+aT)$. Although they are the same at the small $T$ limit, we find empirically that the latter is more accurate. 
With the two changes, we finally obtain the conversion formula:
\begin{equation} \label{eq:bs2n}
\sigman(K) \approx \sigmabs F_0 \sqrt{k}\left(1+\frac{\log^2 k}{24}\right) \Big/ \left(1 + \frac{\sigmabs^2}{24} T \right) \qtext{for} k=\frac{K}{F_0}.
\end{equation}
Figure~\ref{fig:vol_conv} demonstrates the accuracy of the volatility conversions formulas Eqs.~\eqref{eq:grunspan} and \eqref{eq:bs2n}. Even in the extreme test case of $\sigmabs=200\%$, they still closely approximate the true Bachelier volatility. In particular, our approximation in Eq.~\eqref{eq:bs2n} is at the exact values, while Eq.~\eqref{eq:grunspan} shows a slight deviation.

We convert from $\sigman$ to $\sigmabs(K)$ by approximately inverting Eq.~\eqref{eq:bs2n},
\begin{equation} \label{eq:norm2bs}
\sigmabs(K) \approx \frac{\sigman}{F_0\sqrt{k}} \left(1 + \frac{\sigman^2}{24\, k\, F_0^2} \,T\right)
\Big/ \left(1+\frac{\log^2 k}{24}\right) \qtext{for} k=\frac{K}{F_0}.
\end{equation}
This is also consistent with a special case of the HKLW formula~\citep{hagan2002sabr} with $\beta=0$ and $\vov=0$. However, one should use Eq.~\eqref{eq:norm2bs} with caution because the equivalent $\sigmabs(K)$ in fact does not exist for small $K$. Under the Bachelier model, the $K=0$ option has a nonzero time value, whereas the time value under the BS model should be zero regardless of $\sigmabs$ because $F_T\ge 0$. Therefore, $\sigmabs(K)$ should not exist for sufficiently small $K$, and the availability of $\sigmabs(K)$ in Eq.~\eqref{eq:norm2bs} is potentially misleading.
Along the same line or argument, note that Eq.~\eqref{eq:norm2bs} violates \citet{lee2004moment}'s model-free BS volatility bound, $\sqrt{2|\log k|/T}$ as $k\downarrow 0$. Conversely, the equivalent $\sigman(K)$ always exists for $\sigmabs$ at all $K\ge 0$, and Eq.~\eqref{eq:bs2n} does not have a similar issue. 

With the results above, we can convert the DBS volatility to the Bachelier and BS volatilities, respectively:
\begin{align} 
	\sigman(K) &\approx \sigmad D(F_0) \sqrt{k_\disp}\left(1+\frac{\log^2 k_\disp}{24}\right) \Big/ \left(1 + \frac{\beta^2\sigmad^2}{24} T \right) 
	\qtext{for} k_\disp=\frac{D(K)}{D(F_0)} \label{eq:dbs2n},\\
	\sigmabs(K) &\approx \sigmad \frac{D(F_0)}{F_0} \sqrt{\frac{k_\disp}{k}} \; \frac{1+(\log^2 k_\disp)/24}{1+(\log^2 k)/24} \; \frac{1 + \sigmad^2 (D(F_0)/F_0)^2 (k_\disp/ k) T/24}{1 + \beta^2\sigmad^2 T/24}. \label{eq:dbs2bs}
\end{align}
Eq.~\eqref{eq:dbs2n} is an extension of Eq.~\eqref{eq:bs2n}, and we obtain Eq.~\eqref{eq:dbs2bs} by plugging Eq.~\eqref{eq:dbs2n} into Eq.~\eqref{eq:norm2bs}. Eqs.~\eqref{eq:bs2n} and \eqref{eq:norm2bs} are the special cases of the above two formulas for $\beta=1$ and 0, respectively. Both approximations are highly accurate. In fact, we compute the BS volatility skew in Figure~\ref{fig:skew} with Eq.~\eqref{eq:dbs2bs} for DBS and Eq.~\eqref{eq:norm2bs} for the Bachelier model. It is visually indistinguishable from the plot generated with the exact BS skew for the parameter set we tested.

\section{Greeks, hedging, and exchange margin} \label{sec:greeks} \noindent 
This section discusses the Greeks and delta hedging under the Bachelier and DBS models. We first explain the difference in Greeks between the Bachelier and BS models in Section~\ref{ssec:greeks}. Then, we reconcile the difference with backbone and vega-rotated delta in Section~\ref{ssec:backbone}. The backbone also gives insights on the exchange margin difference between the Bachelier and BS models in Section~\ref{ssec:margin}.

\subsection{Greeks} \label{ssec:greeks} \noindent
As with the BS model, the Greeks of the Bachelier model are analytically tractable. Below, we list them without derivation:
\begin{itemize}
	\item Delta (the price sensitivity to the forward asset price):
	$$ \mathcal{D}_\norm = \frac{\partial C_\norm}{\partial F_0} = N(d_\norm) \qtext{and}
	\mathcal{D}_\norm = \frac{\partial P_\norm}{\partial F_0} = N(d_\norm)-1
$$
	\item Gamma (the delta sensitivity to the forward asset price):
	$$ \mathcal{G}_\norm = \frac{\partial^2 C_\norm}{\partial F_0^2} = \frac{n(d_\norm)}{\sigman \sqrt{T}} \quad \text{(same for the put option)}$$
	\item Vega (the price sensitivity to volatility)
	$$ \mathcal{V}_\norm = \frac{\partial C_\norm}{\partial \sigman} = \sqrt{T}\,n(d_\norm) \quad \text{(same for the put option)}$$
	\item Theta (the price sensitivity to the time-to-maturity):
	$$ \Theta_\norm = \frac{\partial C_\norm}{\partial(-T)} = -\frac{\sigman n(d_\norm)}{2\sqrt{T}} \quad \text{(same for the put option)}$$
\end{itemize}
The above Greeks are based on the undiscounted option price in Eq.~\eqref{eq:bach}.\footnote{The theta for the discounted price differs between call and put options.} Delta and gamma are with respect to the forward price $F_0$, but we can obtain those with respect to the spot price $S_0$ easily using the relationship $\partial/\partial S_0 = e^{(r-q)T} \partial/\partial F_0$.

\begin{table}[ht]
	\caption{\label{tab:greeks}
	The option price and Greeks under the Bachelier and DBS models. For the DBS model, 
	$D(F_T) = \beta F_t + (1-\beta)A$.} \vspace{1ex}
	\centering \small
	\begin{tabular}{|c||c|c|} \hline
		Model & Bachelier & Displaced BS \\ \hline\hline
		Stochastic & Arithmetic BM & Geometric BM \\
		differential equation & $dF_t = \sigman\, dW_t$ & $dF_t /D(F_t)= \sigmad\, dW_t$ \\ \hline
		Normalized moneyness & $\displaystyle d_\norm = \frac{F_0-K}{\sigman\sqrt{T}}$ & $\displaystyle d_{1\disp,2\disp} = \frac{\log(D(F_0)/D(K))}{\beta\sigmad\sqrt{T}} \pm \frac12 \beta\sigmad \sqrt{T}$ \\ \hline
		Call option price & $(F_0 - K) N(d_\norm) + \sigman\sqrt{T} \, n(d_\norm)$ & $(D(F_0)\,N(d_{1\disp}) - D(K)\,N(d_{2\disp}))/\beta$ \\ \hline
		Delta ($\partial/\partial F_0$) & $N(d_\norm)$ & $N(d_{1\disp})$ \\ \hline
		Vega ($\partial/\partial \sigma$)& $n(d_\norm)\sqrt{T}$ & $D(F_0)\,n(d_1)\sqrt{T}$ \\ \hline
		Gamma ($\partial^2/\partial F_0^2$) & $n(d_\norm)\;/\;\sigman\sqrt{T}$ & $n(d_{1\disp})\;/\;D(F_0)\sigmad\sqrt{T}$ \\ \hline
		Theta ($-\partial/\partial T$) & $-\sigman\, n(d_\norm)\;/\;2\sqrt{T}$ & $-\sigmad\, D(F_0) n(d_{1\disp})\;/\;2\sqrt{T}$ \\ \hline
	\end{tabular}
\end{table}
\begin{figure}
	\caption{\label{fig:delta} The option delta as a function of the normalized strike price from various models: Bachelier, DBS ($\beta=1/3,\; 2/3$ and $A=F_0$), and BS. We use $F_0=1$ and $T=1$. For a fair comparison, $\sigman$ and $\sigmad$ at each $K$ are calibrated to the BS option price with $\sigmabs=0.5$. The delta difference between the BS and Bachelier models can be as large as 10\%.}
	\begin{center}
		\includegraphics[width=0.55\linewidth]{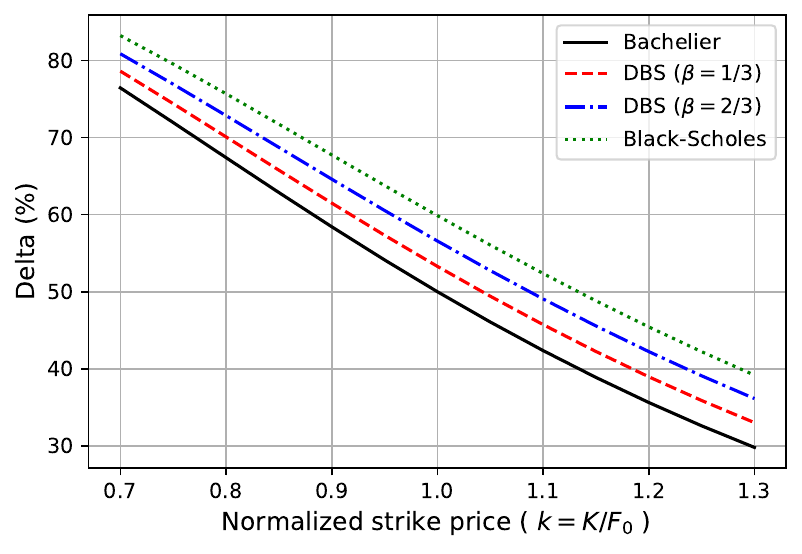}
	\end{center}
\end{figure}

In Table~\ref{tab:greeks}, we show the Greeks under the Bachelier and DBS models side by side for comparison. We adapted the DBS Greeks from the well-known BS Greeks. The notion that the Bachelier model is the $\beta\downarrow 0$ limit of the DBS model also applies to the Greeks. We can reduce the Bachelier Greeks easily from the DBS Greeks from the limit,
$$ D(F_0) \rightarrow A,\quad d_{1\disp}\;\; \text{and}\;\; d_{2\disp} \rightarrow d_\norm, \quad \sigmad A \rightarrow \sigman\; \left(\frac{\partial}{\partial \sigmad}\rightarrow\frac{\partial}{A\,\partial \sigman}\right) \qtext{as}\beta\rightarrow 0.
$$
The Greeks are the partial derivatives with respect to a particular model parameter with the others held constant. Therefore, the Greeks under different models will not be same, even though they measure the sensitivity with respect to the same parameter. Delta is a good example. Figure~\ref{fig:delta} shows the difference in delta across strike prices as the model changes from the Bachelier to DBS ($\beta=1/3$ and $2/3$) and to the BS model. The delta difference can be as large as 10\% between the BS and Bachelier models within the parameter set, resulting in different amounts of hedge.

\subsection{Volatility backbone} \label{ssec:backbone} \noindent
In this section, we explain the source of the delta difference using a concept called the volatility \textit{backbone}. The volatility backbone refers to the observed pattern of the change in the ATM implied volatility as $F_0$ varies. In essence, the leverage effect and the backbone are the same phenomenon; the former is the negative association between the price and volatility observed in the equity market, while the latter is a term coined by fixed-income traders to describe the same association in interest rate dynamics. The backbone also refers to the baseline model under which traders execute their delta hedges. If the interest rate closely follows the Bachelier model (i.e., daily changes are independent of the rate levels), then the market is said to follow a \textit{normal backbone}, and it is optimal for the traders to delta hedge with the Bachelier delta rather than the BS delta. The SABR model gained popularity among traders because they can adjust the volatility backbone by choosing the appropriate $\beta$ parameter. The volatility backbone also has important implications in the risk management of an options portfolio, since efficient delta hedging is critically linked to an accurate measure of the implied volatility dynamics. Furthermore, the calculation of important risk management metrics, such as the value-at-risk (VaR) and the expected shortfall, involve a simulation of the underlying process, and the backbone will determine how the volatility of the underlying process evolve over time. See \citet{neo2019swaption} for a further discussion of the backbone.

\begin{figure} 
	\caption{\label{fig:backbone} The change in the BS volatility skew implied from the Bachelier model with $\sigman=0.5$ when the forward $F_0$ decreases from 1 to $0.9$. The dash-dot (blue) line indicates the change of the ATM BS volatility.}
	\begin{center}
		\includegraphics[width=0.55\linewidth]{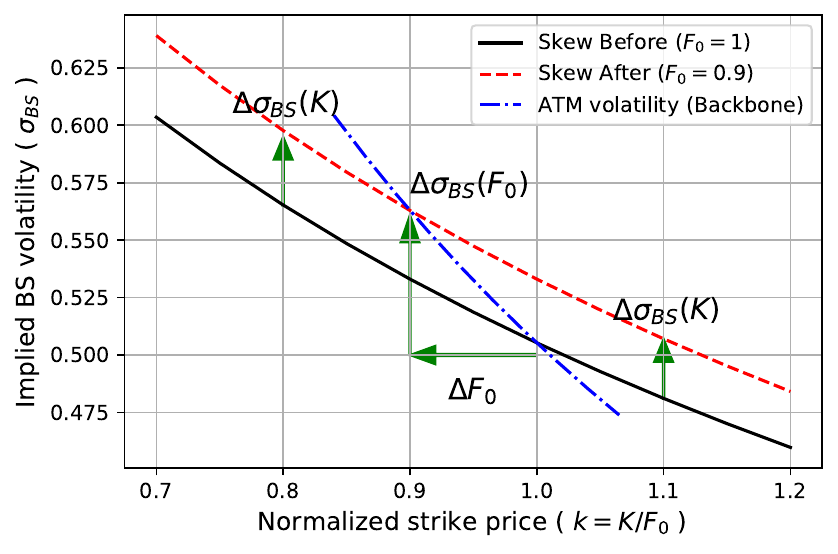}
	\end{center}
\end{figure}

We adjust the delta under the existence of a backbone to 
$$\frac{\partial C}{\partial F_0} = \mathcal{D} + \frac{\partial \sigmabs}{\partial F_0}\mathcal{V},
$$
where the second term is the price change from the induced volatility change and vega (i.e., vega-rotated delta). We use this framework to explain the delta difference between the Bachelier and BS models. Although the Bachelier volatility $\sigman$ is constant, the Bachelier model exhibits a normal backbone when converted to the BS volatility $\sigmabs(K)$. From the leading order term of Eq.~\eqref{eq:norm2bs}, $\sigmabs(K) \approx \sigman/\sqrt{K F_0}$, we can approximate the changes in the BS volatilities ATM and at a fixed $K$ as follows:
$$ \frac{\partial \sigmabs(F_0)}{\partial F_0} \approx -\frac{\sigmabs}{F_0} \qtext{and}
\frac{\partial \sigmabs(K)}{\partial F_0} \approx -\frac{\sigmabs}{2 F_0}.
$$
Note that as $F_0$ decreases, the ATM volatility increases twice as fast as the volatility at $K$ does. Figure~\ref{fig:backbone} illustrates this point. It shows the implied BS skew of the Bachelier model when $F_0$ moves from 1 to 0.9.

We can now express the Bachelier delta by the BS delta with a backbone:
\begin{equation}
	\frac{\partial C_\norm}{\partial F_0} = \mathcal{D}_\norm = \mathcal{D}_\bs + \frac{\partial \sigmabs}{\partial F_0}\mathcal{V}_\bs \approx \mathcal{D}_\bs - \frac{\sigmabs }{2F_0}\mathcal{V}_\bs.
\end{equation}
Therefore, we can understand the delta difference as the vega-rotated delta due to the volatility backbone. We can also obtain the delta difference directly:
$$
\mathcal{D}_\norm - \mathcal{D}_\bs = N(d_\norm) - N(d_1) \approx (d_\norm - d_1) n(d_1) \approx - \frac{\sigmabs\sqrt{T}}{2} n(d_1) = -\frac{\sigmabs}{2F_0}\mathcal{V}_\bs,
$$
where we use the approximation
$$
d_\norm - d_{1\disp} = \frac{(F_0-K)}{\sigman\sqrt{T}} - \frac{\log(F_0/K)}{\sigmabs\sqrt{T}} - \frac12 \sigmabs \sqrt{T} \approx - \frac12 \sigmabs \sqrt{T}.
$$	
The first two terms cancel each other out from the leading-order term of Eq.~\eqref{eq:grunspan}.

Under the DBS model, we can generalize the induced BS volatility change to
$$ \sigmabs(K) \approx \sigmad \frac{\sqrt{D(F_0) D(K)}}{\sqrt{F_0 K}}, \qquad \frac{\partial \sigmabs(K)}{\partial F_0} \approx -\left(1-\frac{\beta F_0}{D(F_0)}\right)\frac{\sigmabs(K)}{2F_0},
$$
where $\beta$ controls the degree of the backbone. Therefore, the DBS model offers a flexible model choice to fit the market-observed backbone. With a single degree of freedom, however, the DBS model cannot fit both the BS volatility skew and the backbone at the same time.

\subsection{Exchange margin} \label{ssec:margin}\noindent
We compare the exchange margin requirement for option positions between the Bachelier and BS models. It is impossible to apply the BS model when the underlying asset price is negative, which is why CME and ICE switched their margin calculation model for oil futures. The margin comparison in this section is valid only when the asset price is still positive, such that the BS implied volatility exists.

\begin{table}
	\caption{\label{tab:span} The price change of an ATM put option ($\sigmabs = 0.5, F_0=K=1$, and $T=1$) under the 16 SPAN risk arrays as percentage of the underlying forward price. We assume a 10\% price scan range and a 25\% volatility scan range. The worst loss (SPAN risk) of the long put option is 9.40\% and 8.63\% under the Bachelier and BS models, respectively, from Scenario 12. The SPAN risk of the short position is 10.26\% and 8.89\%, respectively, from Scenario 13.} \vspace{1ex}
	\centering \small
	\begin{tabular}{|c||c|c|c|c|} \hline 
		& Underlying &  & \multicolumn{2}{c|}{Price change (\%)} \\ \cline{4-5}
		Scenario & asset price move & Volatility move & Bachelier & BS \\
		\hline
		1 / 2 & Unchanged & Up / Down 100\% & 4.94 / $-$4.94 & 4.79 / $-$4.87 \\ \hline
		3 / 4 & Up 33\% & Up / Down 100\% & 3.30 / $-$6.54 & 3.57 / $-$6.23 \\ \hline
		5 / 6 & Down 33\% & Up / Down 100\% & 6.64 / $-$3.21 & 6.08 / $-$3.39 \\ \hline
		7 / 8 & Up 67\% & Up / Down 100\% & 1.75 / $-$8.03 & 2.41 / $-$7.48 \\ \hline
		9 / 10 & Down 67\% & Up / Down 100\% & 8.41 / $-$1.36 & 7.45 / $-$1.79 \\ \hline
		11 / 12 & Up 100\% & Up / Down 100\% & 0.26 / $-$9.40 & 1.31 / $-$8.63 \\ \hline
		13 / 14 & Down 100\% & Up/Down 100\% & 10.26 / 0.60 & 8.89 / -0.06 \\ \hline
		15 / 16 & Up/Down 300\% & Up 100\% & $-$2.41 / 7.59 & $-$1.39 / 6.42 \\ \hline
	\end{tabular}
\end{table}

The margin calculation uses the Standard Portfolio Analysis of Risk (SPAN) developed by \citet{cme2019span}. SPAN uses 16 scenarios called the SPAN risk arrays, and the margin is calculated as the worst potential loss suffered by the portfolio under the risk arrays.\footnote{According to SPAN, scenarios 15 and 16 are for the extreme market moves, and the resulting gain/loss is multiplied by $1/3$.} SPAN first defines the reference amount of the underlying asset price and volatility moves, called the \textit{price scan range} and \textit{volatility scan range}, respectively. The 16 scenarios are then defined as various linear combinations of price scan range and volatility scan range. The specific values of the scan ranges and the 16 scenarios are determined by the exchanges. 

Table~\ref{tab:span} compares the price changes of an ATM put option under the SPAN risk arrays. We use the 16 scenarios from the latest example in \citet{cme2019span} with 10\% price scan range and 25\% volatility scan range. Because the volatility scan range is typically given in percentages, it can be consistently applied to both BS and Bachelier volatilities. In each model, the largest upward price move comes from Scenario 12 (price up and volatility down), while the largest downward move comes from Scenario 13 (price down and volatility up). This is because the put option has a negative delta and positive vega risk. It is important to note that the price moves are larger in the Bachelier model than in the BS model because of the volatility backbone. In the Bachelier model, a downward (upward) move in the underlying asset price induces an upward (downward) move in the implied BS volatility, which is in the same direction as the volatility moves from the risk scenario. This makes the price change larger under the Bachelier model. Consequently, the margin for the long (short) put option is 9.40\% (10.26\%) of the current price under the Bachelier model, which is higher than 8.63\% (8.89\%) under the BS model.

\begin{figure}
	\caption{\label{fig:span} The worst loss (SPAN risk) of a long (left) and short (right) option position under the Bachelier and BS models as functions of strike. We use $\sigmabs = 0.5, F_0=1$, and $T=1$ for the option, and the loss is expressed as a percent of the underlying forward price.
	For a fair comparison, $\sigman$ at each $K$ is calibrated to the corresponding BS option price.} \vspace{1ex}
	\includegraphics[width=0.49\linewidth]{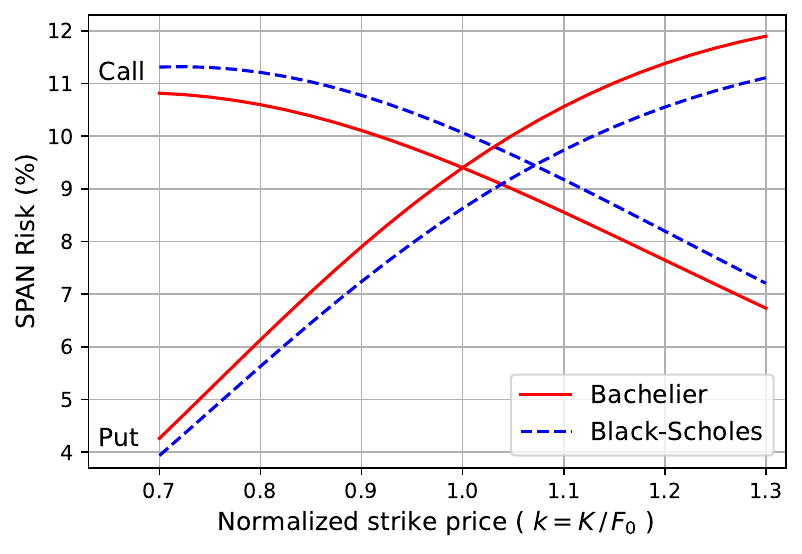}
	\includegraphics[width=0.49\linewidth]{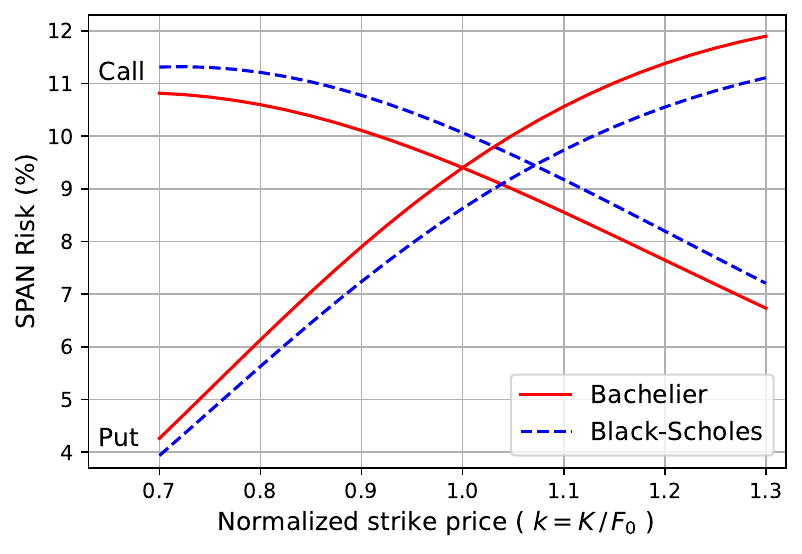}
\end{figure}

For the call option, conversely, the Bachelier model margin is smaller than the BS model margin. Although we do not tabulate the SPAN risk arrays for a call option here in the interest of space, it is straightforward and intuitive to expect that Scenarios 11 (price and volatility up) and 14 (price and volatility down) will cause the largest price moves. In the Bachelier model, the volatility move induced by the price move offsets the volatility move from the risk scenarios. Figure~\ref{fig:span} depicts the margin required for long/shot position of call/put options as functions of strike prices. The relative order of the margin required under the two models remains the same across strike prices, although their difference varies.

\section{Bachelier SV model} \label{sec:sv} \noindent
SV models~\citep{hullwhite1987sv,heston1993closed} have been proposed to explain the volatility smile under the BS model. Similarly, we can extend the Bachelier model to include SV. While research on the Bachelier model with SV is scarce compared to its BS counterpart, we introduce two such models below.

The first model is the SABR model in Eq.~\eqref{eq:sabr}, which provides a Bachelier SV model when $\beta=0$. The equivalent Bachelier volatility is given as a special case of Eq.~\eqref{eq:hagan-new}:
\begin{equation} \label{eq:hagan_n}
	\sigman(K) \approx \sigma_0 H(z)\left(1+\frac{2-3\rho^2}{24}\vov^2 T\right) \qtext{for}
	z = \frac{\vov}{\sigma_0}(K-F_0).
\end{equation}
From this volatility, we can compute the option price with the Bachelier price formula. As we discussed in Section~\ref{ssec:sabr}, this analytical approximation is not restricted by any boundary condition, as $\sigman(K)$ is well defined for negative $K$. This model has been a popular choice in fixed income modeling to handle negative interest rates~\citep{antonov2015free}. Although this approximation is quite accurate, it has some limitations as an analytical approximation. Most importantly, the approximation accuracy deteriorates as $\vov\sqrt{T}$ increases. Moreover, the distribution implied from Eq.~\eqref{eq:hagan_n} is not guaranteed to be arbitrage-free. 

The second model is the hyperbolic normal SV (NSVh) model~\citep{choi2019nsvh}. The NSVh model dynamics are modified from the SABR model to improve 
analytical tractability:
\begin{equation}
	dF_t = \sigma_t \left(\rho\,d\Zdrift[\vov/2]_t + \rhoc\,dX_t\right) \quad \text{and}\quad
	\frac{d\sigma_t}{\sigma_t} = \vov\; d\Zdrift[\vov/2]_t,
\end{equation}
where $Z_t$ and $X_t$ are independent standard BMs, $\Zdrift_t = Z_t + \mu\, t$ denotes BM with drift $\mu$, and $\rhoc = \sqrt{1-\rho^2}$. \citet{choi2019nsvh} shows that the terminal price $F_T$ is distributed as
\begin{equation}\label{eq:FT_nsvh}
\begin{aligned}
	F_T &\distequal\; \mu(F_T) + \frac{\sigma_0}{\vov}\left(\rhoc \sinh\left( \vov W_T + \atanh\, \rho\right) - \rho\, e^{\vov^2 T/2}\right) \\
	&= \mu(F_T) + \frac{\sigma_0}{\vov}\left( \sinh(\vov W_T) + \rho\,\left(\cosh(\vov W_T) - e^{\vov^2 T/2}\right) \right), 
\end{aligned}
\end{equation}
where $\distequal$ denotes the distributional equality and $W_t$ is an independent standard BM. The first equation indicates that the NSVh process follows \citet{johnson1949systems}'s $S_U$ distribution. We can express the vanilla call option price in a closed-form formula\footnote{In the formula, we replace $\mu(F_T)$ in Eq.~\eqref{eq:FT_nsvh} with $F_0$ to maintain consistency with the price formulas in the other models, where $\mu(F_T)=F_0$ holds. Under the NSVh model, $F_t$ is not a martingale and $ \mu(F_T) = F_0 e^{\vov^2 T/2}$ to be exact. The put option price under the NSVh model is
$$\Put{sv}{}(K) = (K - F_0) N(-d_\sv) + \frac{\sigma_0}{2\vov} e^{\vov^2 T/2} \Big((1+\rho)N(d_\sv+\vov\sqrt{T}) - (1-\rho)N(d_\sv-\vov\sqrt{T})-2\rho N(d_\sv)\Big).
$$}:
\begin{equation} \label{eq:jsu_opt}
\begin{gathered}
	\Call{sv}{}(K) = (F_0-K) N(d_\sv) + \frac{\sigma_0}{2\vov} e^{\vov^2 T/2} \left((1+\rho)N(d_\sv+\vov\sqrt{T}) - (1-\rho)N(d_\sv-\vov\sqrt{T})-2\rho N(d_\sv)\right)\\
	\text{for}\quad d_\sv = \frac{1}{\vov\sqrt{T}}\left( \atanh\,\rho + \asinh\left( 
\frac{\vov(F_0-K)}{\rhoc\sigma_0} - \frac{\rho}{\rhoc} e^{\vov^2 T/2} 
\right) \right).
\end{gathered}
\end{equation}
Thanks to its analytical tractability, the option prices under the NSVh model are arbitrage-free. Eq.~\eqref{eq:jsu_opt} also converges to Eq.~\eqref{eq:bach} as $\vov\downarrow 0$.

\begin{figure}
	\caption{\label{fig:nsvh} The Bachelier volatility smile generated by the NSVh~\citep{choi2019nsvh} (left column) and SABR models with $\beta=0$ (right column). From the base parameters, $F_0=100$, $\sigman(F_0)=20$, $\nu=0.2$, $\rho=0.1$, we vary $\nu$ (top row) and $\rho$ (bottom row) to illustrate that the vol-of-vol, $\vov$, and the correlation, $\rho$, control the convexity and slope of the volatility smile.} \vspace{1ex}
	\includegraphics[width=0.49\linewidth]{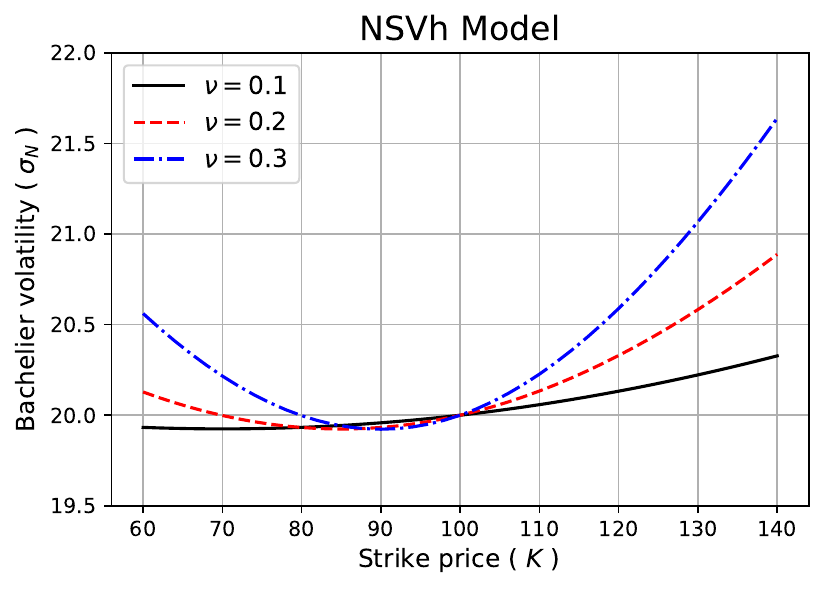}
	\includegraphics[width=0.49\linewidth]{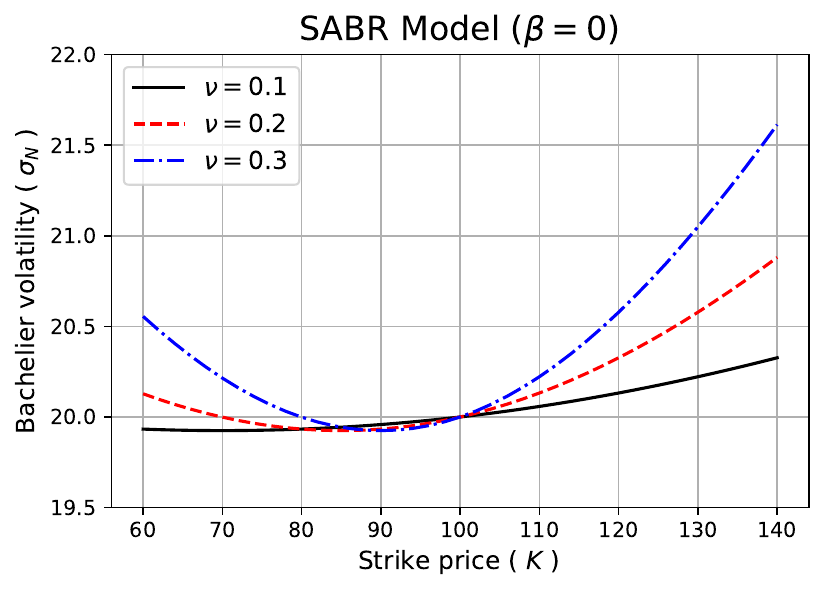}\\
	\includegraphics[width=0.49\linewidth]{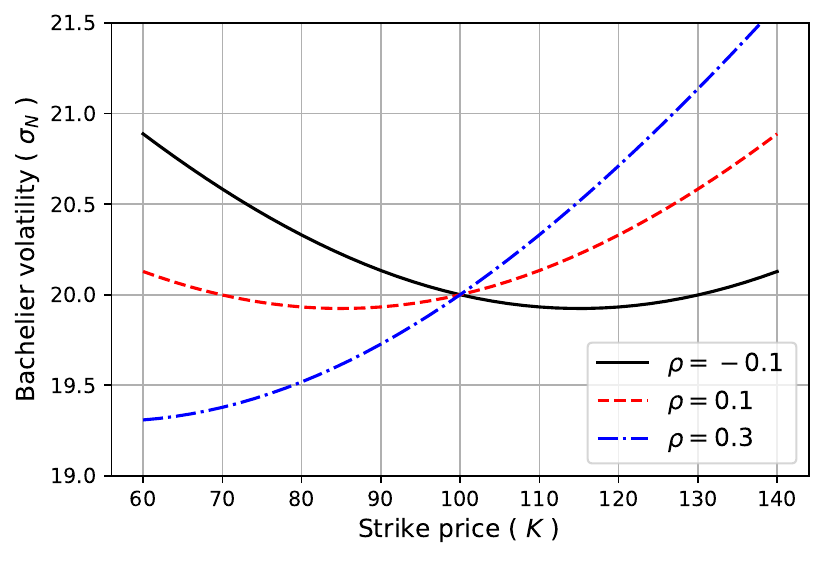}
	\includegraphics[width=0.49\linewidth]{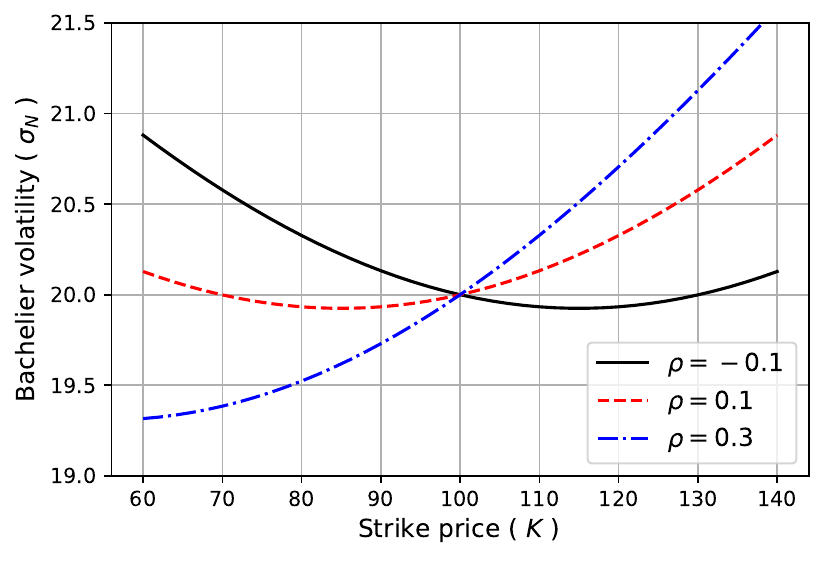}
\end{figure}

As the SABR ($\beta=0$) and NSVh models share the same root, the parameters $\sigma_0$, $\vov$, and $\rho$ have similar effects on the volatility smile a in the two models: $\sigma_0$ controls the level of the smile, $\vov$ the convexity, and $\rho$ the slope. With three degrees of freedom, we can calibrate the models to the observed volatility smile. If we observe the ATM volatility, $\sigman(F_0)$, from the market, then the calibration becomes simpler because we can solve for $\sigma_0$ that yields $\sigman(F_0)$. We can adjust Eqs.~\eqref{eq:hagan_n} and \eqref{eq:jsu_opt}, respectively, to 
\begin{equation} \label{eq:hagan_n2}
	\sigman(K) = \sigman(F_0) H(z)
\end{equation}
and
\begin{equation} \label{eq:jsu_opt2}
\begin{gathered}
	\Call{sv}{}(K) = (F_0-K) N(d_\sv) + \sigman(F_0)\sqrt{\frac{T}{2\pi}} \frac{(1+\rho)N(d_\sv+\vov\sqrt{T}) - (1-\rho)N(d_\sv-\vov\sqrt{T})-2\rho N(d_\sv)}{(1+\rho)N(d_\sv^0 + \vov\sqrt{T}) - (1-\rho)N(d_\sv^0 - \vov\sqrt{T})-2\rho N(d_\sv^0)} \\
\text{for}\quad d_\sv^0 = d_\sv \big|_{K=F_0} =\frac{1}{\vov\sqrt{T}}\left( \atanh\,\rho - \asinh\left(\frac{\rho}{\rhoc} e^{\vov^2 T/2} \right) \right).
\end{gathered}
\end{equation}
This is possible under the NSVh model because the ATM option price is proportional to $\sigma_0$. 

In Figure~\ref{fig:nsvh}, we show the Bachelier volatility smile generated by the NSVh and SABR ($\beta=0$) models for varying $\vov$ and $\rho$ but with a fixed $\sigman(F_0)$. When the two models are calibrated to the same ATM volatility, the volatility smiles are very close to each other. See \citet{choi2019nsvh} for further numerical evidence supporting the equivalence between the two models.

Besides the two models we introduced above, \citet{sun2016implied} also explores the stochastic Bachelier volatility model by presenting parameterized forms of the volatility smile. \citet{perederiy2018vannavolga} extends the Vanna-Volga method~\citep{castagna2007option} from the original BS volatility context to the Bachelier model, which is helpful for the arbitrage-free interpolation of the volatility smile. Finally, while not strictly speaking an SV model, \citet{karami2018appl} provide an asymptotic expansion method to obtain the equivalent Bachelier volatility of the general local volatility models.

\section{Pricing other derivatives}\label{sec:exotic} \noindent 
In this section, we derive pricing formulas for the two types of exotic claims under the Bachelier model. The exotic claims covered in this section are typically traded in over-the-counter markets and therefore are not subject to the model switch at CME or ICE. Moreover, accurate pricing of these claims requires more complicated models that can be calibrated to the current and future volatility skew. In Section~\ref{ssec:barrier}, we use the DBS model to price the barrier option. Although $\beta$ provides limited freedom to fit the skew, the model is far from being used as a pricing model in practice. The Bachelier SV model in Section~\ref{sec:sv} has the potential for pricing exotic derivatives, but a discussion of this point goes beyond the scope of this paper. For further information on barrier options pricing under the replication approach, we refer to \citet{derman1995static}. Therefore, the purpose of this section is to provide a complete reference to the Bachelier model.

\subsection{Basket, spread, and Asian options} \noindent
Basket options, spread options, and Asian options are options with payouts that depend on the linear combination of multiple correlated asset prices.
Under the BS model, it is very difficult to derive the exact pricing of these claims because the linear combination of log-normal random variables is no longer log-normally distributed. Therefore, pricing such claims under the BS model requires either a simplifying approximation or numerical schemes. See \citet{choi2018sumbsm} and the references therein for a review of such methods. Under the Bachelier model, however, the pricing becomes trivial because the weighted sum of the correlated arithmetic BMs remains normally distributed. 

We first consider basket and spread options. Let $N$ assets follow the correlated Bachelier model:
$$ dF_{t,k} = \sigma_{\norm, k}\, dW_k, \quad dW_i dW_j = \rho_{ij} dt \;\; (\rho_{ii}=1).$$
Let us define $\Sigma$ as the covariance matrix of the terminal asset prices, $F_{T,k}\; (k = 1\ldots N)$, whose $(i,j)$ element is given by $\Sigma_{ij} = \rho_{ij} \sigma_{\norm, i} \sigma_{\norm, j}\, T$.
Suppose that the basket portfolio consists of $N$ assets with weight vector $w$; that is,
$$ B_t = \sum_{k=1}^{n} w_k F_{t,k}, $$
and that the call option payout is $\max(B_T - K,\, 0)$ for the strike price $K$. Then, the mean and standard deviation of the portfolio value $B_T$ are, respectively,
$$ \mu(B_T) = B_0 \qtext{and} sd(B_T) = \sqrt{w^\top \Sigma w}.
$$
We can then calculate the price of the basket option using the generalized Bachelier formula in Eq.~\eqref{eq:bach_gen}. The spread option is a special case of the basket option with $N=2$ and $w^\top=(1,-1)$:
$$ \mu(B_T) = F_{0,1} - F_{0,2} \qtext{and} sd(B_T) = \sqrt{(\sigma^2_{\norm,1} - \rho_{12}\sigma_{\norm,1}\sigma_{\norm,2} + \sigma^2_{\norm,2})T}.
$$
The Bachelier price of the spread option with $\sigma_{\norm, k} \approx F_{0,k} \sigma_{\bs, k}$ serves as an approximation of the BS price~\citep{poitras1998spread}.

Asian options use the average asset price over time for the payout. We can also consider Asian options as a type of basket option where the payout is a linear combination of the prices of a single asset at different times. In the case of discretely monitored Asian options, the average price is
$$ A_T = \frac{1}{N}\sum_{k=1}^N F_{t_k} \qtext{for} 0\le t_1 < \cdots < t_N = T.$$
Since the covariance between the two observations, $F_{t_i}$ and $F_{t_j}$ is 
$$ \Sigma_{ij} = \sigman^2 \min(t_i, t_j), $$
we can the express the Asian options prices within the same pricing framework as the basket options formulated above.

In the case of the continuously monitored Asian option, the price is continuously averaged between time $S$ and $T$:
$$A_T = \frac{1}{T-S}\int_{t=S}^{T} F_t\, dt = \frac{\sigman}{T-S} \int_{t=S}^{T} W_t\, dt \quad (S<T).$$
From the property of BM, it is not difficult to show that the variance of $A_T$ is
$$ \text{Var} (A_T) = \sigman^2 \left(\frac{2S + T}{3}\right). 
$$
Therefore, we can derive the price of an Asian option under the Bachelier model using the generalized Bachelier formula in Eq.~\eqref{eq:bach_gen} with 
$$ \mu(A_T) = F_0 \qtext{and} sd(A_T) = \sigman \sqrt{\frac{2S + T}{3}}.
$$

\subsection{Barrier options} \label{ssec:barrier} \noindent
Next, we derive the barrier options pricing formulas under the Bachelier model. Although the derivation does not entail mathematical difficulty, we offer the first derivation of the pricing formulas to the best of our knowledge. The barrier option price under the BS model is analytically available~\citep{haug2007complete}. We will again show that the BS barrier price converges to the Bachelier price in the $\beta\downarrow 0$ limit of the DBS model.

We consider the following four types of knock-out barrier options:
\begin{itemize}
\item \textbf{Down-and-out call option} with strike price $K$ and knock-out barrier $L$ whose price is denoted by $\Call{n}{do}(K, F_0; L)$.
\item \textbf{Up-and-out call option} with strike price $K$ and knock-out barrier $H$ ($> K$) whose price is denoted by $\Call{n}{uo}(K, F_0; H)$.
\item \textbf{Down-and-out put option} with strike price $K$ and knock-out barrier $L$ ($< K$) whose price is denoted by $\Put{n}{do}(K, F_0; L)$.
\item \textbf{Up-and-out put option} with strike price $K$ and knock-out barrier $H$ whose price is denoted by $\Put{n}{uo}(K, F_0; H)$.
\end{itemize}
By the nature of the knock-out option, we assume that $L<F_0<H$; otherwise, the option is already knocked out and the $t=0$ price should be zero. The second (up-and-out call) and third (down-and-out put) options have extra restrictions on the barrier, $K<H$ and $L<K$, respectively.
Without these conditions, the option is worthless because the path of $F_t$ always triggers the barrier before it reaches the in-the-money payout region. Note that the list of barrier options above is exhaustive because we can compute the corresponding knock-in options price through the so-called 
in-and-out parity. 

We define the running maximum and minimum of $F_t$ during the period $[0,t]$ as $F_T^M=\max \limits_{0\le t \le T}F_t$ and $F_T^m=\min \limits_{0\le t \le T}F_t$, respectively. We can find the PDF of $F_T$ conditional on $F_T^M$ and $F_T^m$, respectively, using the reflection principle~\citep[\S~1.8]{harrison1985brownian}:
\begin{align*}
\mathbb{P}(F_T-F_0\in dx, F^M_T-F_0 < y) &= f(x,y)\, dx \quad (x\leq y,\; 0 \leq y),\\
\mathbb{P}(F_T-F_0\in dx, F^m_T-F_0 > y) &= f(x,y)\, dx \quad (x\leq y,\; y \leq 0),
\end{align*}
where
$$ f(x,y) = \frac{1}{\sigman\sqrt{T}}\left( n\!\left(\frac{x}{\sigman\sqrt{T}}\right)- n\!\left(\frac{x-2y}{\sigman\sqrt{T}}\right) \right).
$$
Based on these conditional PDFs, we can express the barrier option prices as 
\begin{gather*}
\Call{n}{do} = \int_K^{\infty} (x-K) f(x-F_0,L-F_0)\, dx, \quad \Call{n}{uo} = \int_{K}^{H} (x-K) f(x-F_0,H-F_0)\, dx \\
\Put{n}{do} = \int_{L}^{K} (K-x) f(x-F_0,L-F_0)\, dx, \quad \Put{n}{uo} = \int_{-\infty}^{K}(K-x) f(x-F_0,H-F_0)\, dx.
\end{gather*}

We express the outcome succinctly by taking advantage of the vanilla option price with suboptimal exercise policy. For a call (put) option struck at $K$, suppose that the option holder exercises it when $F_T>K^*$ ($F_T<K^*$) for some $K^*$. The call option value under the suboptimal exercise, $C_\norm(K, F_0; K^*)$, is \footnote{
The suboptimal put option price is
$$P_\norm(K, F_0; K^*) = (K - F_0) N(-d^*_\norm) + \sigman\sqrt{T} \, n(d^*_\norm)$$}
\begin{equation} \label{eq:bach_h}
C_\norm(K, F_0; K^*) = (F_0 - K) N(d^*_\norm) + \sigman\sqrt{T} \, n(d^*_\norm)
\qtext{for} d^*_\norm = \frac{F_0 - K^*}{\sigman \sqrt{T}}.
\end{equation}
This value is always less than the regular price, $C_\norm(K, F_0)$ if $K^*\neq K$ and is equal to $C(K, F_0)$ only if $K^*= K$. Using the suboptimal price expressions, the barrier option prices under the Bachelier model are conveniently given by
\begin{align}
\Call{n}{do}(K, F_0; L) &= C_\norm(K, F_0)- C_\norm(K, 2L-F_0) \label{eq:DO_call}\\
\Call{n}{uo}(K, F_0; H) &= C_\norm(K, F_0) - C_\norm(K, F_0; H) - C_\norm(K, 2H-F_0) + C_\norm(K, 2H-F_0; H) \label{eq:UO_call} \\
\Put{n}{do}(K, F_0; L) &= P_\norm(K, F_0)-P_\norm(K, F_0; L) - P_\norm(K, 2L-F_0) + P_\norm(K, 2L-F_0; L) \label{eq:DO_put} \\
\Put{n}{uo}(K, F_0; H) &= P_\norm(K, F_0) - P_\norm(K, 2H-F_0). \label{eq:UO_put}
\end{align}
In \ref{apdx:bs_barrier}, we also present the corresponding barrier option prices under the BS and DBS models. 

\begin{figure} 
	\caption{\label{fig:barrier} Price of the knock-out option as a function of the barrier price for various models: Bachelier, DBS ($\beta=1/3, 2/3$ with $A=F_0$), and BS. We use $F_0=K=1$, $T=1$, and calibrate the implied volatility to the ATM option price of 0.2 ($\sigman\approx \sigmad\approx\sigmabs\approx0.5$). We display the price of a down-and-out put option for $L<1$ and up-and-out call option for $H>1$.}
	\begin{center}
		\includegraphics[width=0.55\linewidth]{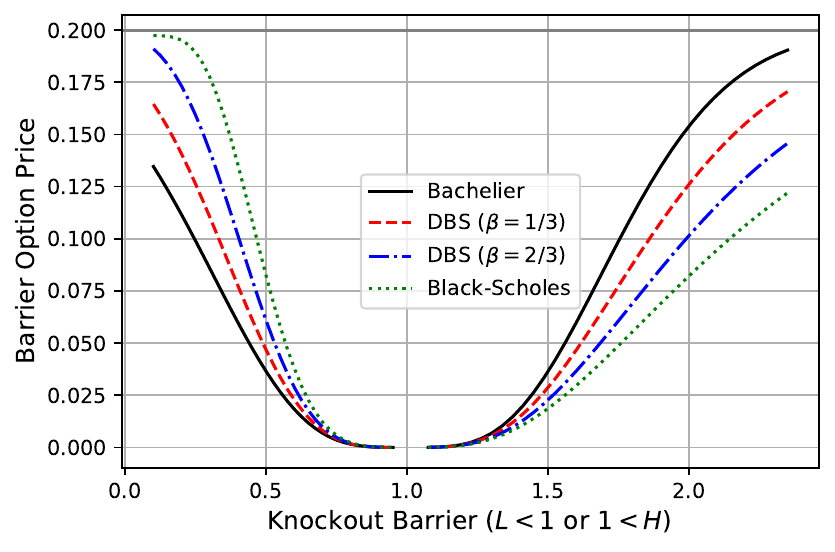}
	\end{center}
\end{figure}

In Figure~\ref{fig:barrier}, we depict the price of the knock-out option with $K=F_0$ as a function of the barrier price, $L$ or $H$, for different models. As expected, the prices from the Bachelier and BS models are the two end points of the price spectrum. 
We also observe the impact of the volatility skew on the barrier option price in Figure~\ref{fig:barrier}. Figure~\ref{fig:skew} illustrates the BS volatility skew generated by the same set of models. The barrier option price depends on the volatilities at the barrier and the strike price. When the volatility at the strike (and consequently the vanilla option price at the strike) is the same, the knock-out option price decreases as the volatility at the barrier increases because the knock-out probability increases. The down-and-out put price ($L<1$ in Figure~\ref{fig:barrier}) indeed decreases as $\beta\downarrow 0$ due to the increase in the implied BS volatility at low strikes, as Figure~\ref{fig:skew} shows. Conversely, the up-and-out call option price ($H>1$) increases as $\beta\downarrow 0$ because the BS volatility at high strikes moves in the opposite direction. Therefore, to price the barrier option correctly, one should use the DBS model with the right $\beta$ parameter that fits the market volatility skew. Following this line of argument, we note that the Bachelier and BS models are just two possible model choices within the DBS model family.

\section{Conclusion} \label{sec:conc} \noindent
Bachelier proposed the very first option pricing model that predates the BS model by more than 70 years. Over time, however, it was eclipsed by the BS model, as academics and practitioners alike expected that asset prices would be strictly positive. The negative prices of the oil futures contracts at the CME, and the subsequent urgent model switch, put the Bachelier model back under spotlight. In fact, prior to this, the fixed income market already switched to quoting implied Bachelier volatilities as Euro and CHF rates became negative in the mid-2010s. 

Interest in the Bachelier model has hitherto been historical in nature. Some review papers cover the life of Louis Bachelier, and some studies compare the Bachelier and BS models in terms of vanilla option pricing formulas. Unlike the BS model counterpart, which has a rich literature covering the pricing of various liquid exotic options, the Bachelier model lacks a similar exposition in the literature.
As the CME and ICE changed their models for oil futures derivatives to the Bachelier model, the model drew more attention and will be adopted for a broader range of financial products. In line with this change, we provide a comprehensive review of various topics related to the Bachelier model for both researchers and practitioners. Specifically, we cover topics such as implied volatility inversion, volatility conversion between related models, Greeks and hedging for risk management, SV models, and the pricing of exotic options such as basket, spread, Asian, and barrier options. We also connect the Bachelier and BS models by introducing the DBS model, and thus offer a continuous spectrum of model choices between the two models. We place the Bachelier model in the option pricing literature by showing its connection to the other mainstream pricing models, notably the CEV and SABR models. With this paper, we hope to see the Bachelier model receive more attention in both research and application, and this paper is intended as a one-stop reference for academics and practitioners already familiar with the BS model and exploring the use of the Bachelier model.

\section*{Acknowledgments}
The authors are grateful to the journal's editors, Robert Webb and Bart Frijns, and two anonymous reviewers for valuable comments and suggestions.
Jaehyuk Choi was supported by the Bridge Trust Asset Management Research Fund.
Minsuk Kwak acknowledges that this work was supported by Hankuk University of Foreign Studies Research Fund of 2021. Minsuk Kwak also acknowledges that this work was supported by the National Research Foundation of Korea(NRF) grant funded by the Korea government(MSIT) (NRF-2019R1F1A1062885).

\section*{Data Availability Statement} \noindent
The python implementations that produce the findings of this study are openly available at \url{https://github.com/PyFE/PyfengForPapers}.

\appendix
\section{Barrier option prices under the (Displaced) BS model} \label{apdx:bs_barrier}
For the barrier option prices under the BS model, we refer to \citet{zhang2001exotic} and \citet{haug2007complete}. Similar to Eq.~\eqref{eq:bach_h}, we define the suboptimal value of the call option where the holder incorrectly exercises the option when $F_T\ge K^*$: 
\begin{equation} \label{eq:bsm_h}
C_\bs(K, F_0;K^*) = F_0 N(d^*_1) - K N(d^*_2) \qtext{for}
d^*_{1,2}=\frac{\text{log}(F_0/K^*)}{\sigmabs \sqrt{T}} \pm \frac{\sigmabs \sqrt{T}}{2}.
\end{equation}
This value is equal to the regular option value, $C_\bs(K, F_0)$, only when $K=K^*$. With the suboptimal option prices, we can express the BS barrier option as~\citep{haug2007complete}:
\begin{align}
\Call{bs}{do}(K, F_0; L) &= C_\bs (K, F_0) - \frac{F_0}{L}\, C_\bs\left(K, \frac{L^2}{F_0}\right) \\ 
\Call{bs}{uo}(K, F_0; H) &= C_\bs (K, F_0) - C_\bs (K, F_0; H) - \frac{F_0}{H} \left(C_\bs\left(K, \frac{H^2}{F_0}\right) - C_\bs\left(K, \frac{H^2}{F_0};H\right)\right) \\
\Put{bs}{do}(K, F_0; L) &= P_\bs (K, F_0)-P_\bs (K, F_0; L) - \frac{F_0}{L} \left(P_\bs\left(K, \frac{L^2}{F_0}\right)-P_\bs\left(K, \frac{L^2}{F_0}; L\right) \right) \\
\Put{bs}{uo}(K, F_0; H) &= P_\bs (K, F_0) - \frac{F_0}{H}\, P_\bs\left(K, \frac{H^2}{F_0}\right),
\end{align}
where we assume that $L < F_0 < H $.

We can obtain the barrier option price under the DBS model by making the following substitutions:
$$ \sigmabs \rightarrow \beta \sigmad, \quad K \rightarrow D(K), \quad F_0 \rightarrow D(F_0),\quad L \rightarrow D(L),\quad H \rightarrow D(H)
$$
and dividing the final result by $\beta$. For example, the down-and-out call option price is
$$ \Call{d}{do}(K, F_0; L) = \frac{C_\bs (D(K), D(F_0))}{\beta} - \frac{D(F_0)}{\beta\,D(L)}\, C_\bs\left(D(K), \frac{D(L)^2}{D(F_0)}\right).
$$

\singlespacing
\bibliography{@Bib/NormalModel_Z2,@Bib/SV_Z2}

\end{document}